\documentclass[pre,aps,12pt]{revtex4-2}
\usepackage{amsmath}
\usepackage{amssymb}
\usepackage{graphicx}

\begin{document}

\title{Spin-$s$ model with competing interactions on diamond-decorated lattices}
\author{D.~V.~Dmitriev}
\email{dmitriev@deom.chph.ras.ru}
\author{V.~Ya.~Krivnov}
\author{O.~A.~Vasilyev}
\affiliation{Institute of Biochemical Physics of RAS, Kosygin str.
4, 119334, Moscow, Russia.}
\date{}

\begin{abstract}
We investigate the ground state properties, magnetization, and
low-temperature thermodynamics of the
ferromagnetic-antiferromagnetic spin-$s$ model on
diamond-decorated lattices with ideal diamond units, incorporating
bilinear Heisenberg and higher-order exchange interactions between
diagonal spins-$\sigma$. Local conservation of the
composite spin on each diamond diagonal enables exact analysis. For the pure Heisenberg case,
the system undergoes a series of $2\sigma$ transitions between
monomer-dimer (MD), ferrimagnetic (Ferri) and ferromagnetic (F) phases with
different optimal composite spin values as the coupling ratio
varies. In the presence of higher-order interactions, a
multicritical point exists where the states with all possible
values of composite spin are degenerate, leading to maximal ground
state degeneracy. The case $s=\sigma=1$ with
bilinear and biquadratic interactions is studied in detail. Its
phase diagram comprises three phases - F, Ferri and MD, which meet at a triple point.
On the phase boundaries, the ground state becomes macroscopically
degenerate. For the diamond chain, we calculate the ground state
degeneracy exactly; for higher dimensions, the problem maps onto a
bond percolation framework, solved numerically. The residual
entropy per spin reaches up to $60\%$ of the maximal value,
peaking at the triple point. Low-temperature magnetization curves in external magnetic fields exhibit plateaus and jumps. The excitation spectrum is gapped in the MD phase, gapless in the F phase, and resembles that of the Lieb-Mattis ferrimagnet in the Ferri phase. The high
residual entropy suggests potential applications in
ultra-low-temperature cooling and quantum thermal machines.
\end{abstract}

\maketitle

\section{Introduction}

Quantum magnets on geometrically frustrated lattices have been
extensively studied over the past decades due to their ability to
host exotic ground states, macroscopic degeneracies, and
unconventional thermodynamic properties \cite{diep, diep2, mila}.
Among the most intriguing phenomena arising from frustration is
the emergence of dispersionless (flat) magnon bands, where magnons
become localized within specific trapping cells as a result of
destructive quantum interference \cite{schulenburg2002macroscopic,
flat, derzhko2007universal}. This flat-band physics has been
observed in a broad class of highly frustrated antiferromagnetic
spin-$\frac{1}{2}$ systems, including the kagome and pyrochlore
lattices \cite{zhitomirsky2005high}. In antiferromagnetic
flat-band models, localized states constitute exponentially
degenerate ground state manifolds in the saturation magnetic
field, leading to distinctive thermodynamic features such as
magnetization plateaus, low-temperature anomalies in the specific
heat, and enhanced magnetocaloric effects \cite{shulen, schmidt,
honecker, richter2018thermodynamic, zhitomirsky2003enhanced}.

A distinct route to macroscopic degeneracy emerges in
spin-$\frac{1}{2}$ systems with competing ferromagnetic (F) and
antiferromagnetic (AF) interactions. In these F-AF models, the
zero-temperature phase diagram features multiple phases separated
by quantum critical points. At these critical parameter values,
the ground state becomes macroscopically degenerate in zero
magnetic field, often with a higher residual entropy than in
purely antiferromagnetic flat-band models \cite{zhitomir, Derzhko,
KDNDR, DKRS, anis, anis2}. Prominent examples include the F-AF
delta-chain and its two-dimensional generalizations on Tasaki and
kagome lattices \cite{anis3}.

This macroscopic ground state degeneracy is not only of
fundamental interest but also has practical implications.
Materials with a large density of low-lying energy states are
prime candidates for adiabatic demagnetization cooling, where the
large entropy reservoir enables efficient refrigeration to
millikelvin temperatures. More recently, such systems have been
recognized as promising platforms for quantum thermal machines.
When a thermodynamic cycle operates across a quantum critical
point, the large ground-state degeneracy can function as an
`entropic lever', directly linking machine performance to the
extensive degeneracy of the model \cite{PhysRevE.96.022143,
purkait2022performance, castorene2024effects}.

Attention has recently turned to frustrated spin systems built
from diamond-shaped clusters. A diamond unit, comprising two
monomer spins connected via two bonded spins on the diagonal, can
be either "ideal" (with two distinct exchange interactions) or
"distorted" (with three). The spin-$\frac{1}{2}$ Heisenberg models
composed of ideal diamond units with purely antiferromagnetic
interactions have been intensively studied \cite{Takano,
morita2016exact, hirose2016exact, caci2023phases,
karl2024thermodynamic}, revealing a rich variety of ground state
phases, including the Lieb-Mattis ferrimagnet, a monomer-dimer
phase, and a dimer-tetramer phase. Notably, the latter two exhibit
macroscopic ground state degeneracy. However, an alternative route
to degeneracy exists in F-AF models with ideal diamonds. When the
local energies of states with different values of the composite
spin formed on diamond diagonals coincide, the ground state can be
expressed as a product of eigenstates of isolated ferromagnetic
clusters embedded in a background of diamonds with diagonal
singlets \cite{diamond1d, DKV2D, DKV_Richter}. This scenario,
realized for specific relations between F and AF interactions,
yields exponential growth in the total number of ground states and
high residual entropies.

Experimentally, diamond-based frustrated magnets have attracted
considerable interest. Several recently synthesized compounds
exhibit competing ferromagnetic and antiferromagnetic interactions
within spin-$\frac{1}{2}$ diamond units, including
$K_{3}Cu_{3}AlO_{2}(SO_{4})_{4}$ (alumoklyuchevskite) \cite{alum},
$K_{2}Cu_{3}(MoO_{4})_{4}$ \cite{PhysRevB.111.144420}, and $Fe-Cu$
bimetallic compounds with diamond-decorated honeycomb structures
\cite{HONG20043271}. In addition to these compounds, a diamond
chain with $s=1$ has been identified in the nickel-based polymer
coordination compound
$[Ni_{3}(OH)_{2}(C_{4}H_{2}O_{4})(H_{2}O_{4})_{4}]\cdot 2H_{2}O$
\cite{B207133A}. Compounds with a diamond chain structure
 with higher spin magnitude have been identified in the
cobalt-based compound $Co_{3}(OH)_{2}(C_{4}O_{4})_{2}\cdot
3H_{2}O$ (diamond chain with spin-$\frac{3}{2}$) \cite{mole2011two},
and in the iron-based mixed-valent compound
$[(CH_{3})_{2}NH_{2})]_{6}[Fe^{III}_{4}Fe^{II}_{2}(\mu_3-O)_2
(\mu_3-OH)_2(\mu_3-SO_{4})_{8}]$ (diamond chain with mixed spin
$(\frac{5}{2} ,2)$ ) \cite{sorolla2019mixed}. These materials
provide potential platforms for realizing the rich physics
predicted in theoretical models.

While extensive work has focused on spin-$\frac{1}{2}$ systems
consisting of ideal diamond units, higher-spin analogues remain
comparatively unexplored. A number of interesting results have
been obtained for the $s=1$ antiferromagnetic Heisenberg ideal
diamond chain. It was shown \cite{JPSJ.86.033707} that the ground
state phase diagram of this model consists of the monomer-dimer,
Haldane, and two ferrimagnetic phases. In a non-zero magnetic
field, spin-liquid phases, a bound-magnon crystal, and the
ferromagnetic phase appear in the ground state phase diagram
\cite{Zoshki_2026}. The mixed spin-$(\frac{1}{2},1)$
Ising-Heisenberg diamond chain with bilinear and biquadratic
interactions between the diagonal spin-$1$ sites has been exactly
solved, revealing a rich ground state phase diagram
\cite{rojas2011exactly}.

In this paper, we study the general case of the F-AF Heisenberg
model on diamond-decorated lattices with ideal diamond units for
arbitrary spin values $s$ (monomer spins) and $\sigma$ (diagonal
spins). The local Hamiltonian includes bilinear and arbitrary
higher-order exchange interactions between the diagonal spins,
with the composite spin on each diagonal taking values
$L=0,1,\ldots ,2\sigma$. The inclusion of these interactions adds
new dimensions to the problem. While real diamond compounds with
biquadratic interactions are not yet widely available, the study
of the exotic ground states initiated by these interactions is of
great theoretical interest. The bilinear Heisenberg term
originates from the superexchange process, while higher-order
terms arise from virtual electron hopping and, although typically
smaller in magnitude, can induce novel properties beyond the
standard bilinear Hamiltonian \cite{harada2002quadrupolar,
mila2000origin}. The bilinear-biquadratic model describes spin-$1$
magnetic systems such as the triangular lattice layered materials
$NiGa_{2}S_{4}$ \cite{science.1114727}, $Ba_{3}NiSb_{2}O_{9}$
\cite{PRL.107.197204, PRB.95.060402}, and others. The competition
between these couplings gives rise to rich physics in frustrated
spin systems. Higher-order interactions play a key role, for
instance, in stabilizing the ferromagnetic state in monolayer
$NiCl_{2}$ \cite{ni2021giant}. With the synthesis of novel classes
of magnetic materials, the study of models with higher-order
exchange interactions has become increasingly relevant and timely
\cite{kartsev2020biquadratic}.

As will be demonstrated in this work, the higher-order couplings
significantly alter the system properties and can lead to a
dramatic increase in ground state degeneracy, substantially
enhancing the residual entropy compared to the purely bilinear
case. We first develop the general formalism for arbitrary $s$ and
$\sigma$, showing that the ground state phase diagram is
determined by the local energy of a single diamond. The simplest
nontrivial case, $s=\sigma=1$ with bilinear and biquadratic
interactions, is then studied in detail as a concrete illustration
of the general approach. For this case, we construct the full
ground state phase diagram, determine the macroscopic degeneracies
on the phase boundaries, and analyze the magnetization behavior
and low-temperature thermodynamics. By mapping the degeneracy
problem to a bond percolation framework, we elucidate the role of
lattice geometry in determining residual entropy and magnetic
order. Our results reveal a rich interplay between spin magnitude,
frustration, and interaction parameters, and highlight the
potential of these systems for applications in
ultra-low-temperature refrigeration and quantum thermal machines.

The remainder of this paper is organized as follows. In Section
II, we develop the general formalism for arbitrary spins $s$ and
$\sigma$, derive the ground state phase diagram for the pure
Heisenberg case, discuss the general case with higher-order
interactions, and then illustrate the results explicitly for the
spin-$1$ model with bilinear and biquadratic couplings. Section
III provides an analytical calculation of the ground state
degeneracy on phase boundaries and at the multicritical point for
the diamond chain, first for general $s$ and $\sigma$ and then
specialized to $s=\sigma=1$. In Section IV, we focus on the spin-
$1$ case and extend these calculations to two- and
three-dimensional diamond-decorated lattices by mapping the ground
state degeneracy problem onto a bond percolation framework, which
we solve numerically using a Monte Carlo approach. Sections V and
VI explore the low-temperature magnetization behavior in an
external magnetic field, analyze the one-magnon excitation
spectrum, and discuss the resulting specific heat characteristics
across different phases and critical lines. Finally, Section VI
summarizes our key findings and discusses their implications for
frustrated quantum magnets and potential applications in cryogenic
cooling technologies.

\section{F-AF spin-$s$ model on ideal diamond-decorated lattices}

\begin{figure}[tbp]
\includegraphics[width=4in,angle=0]{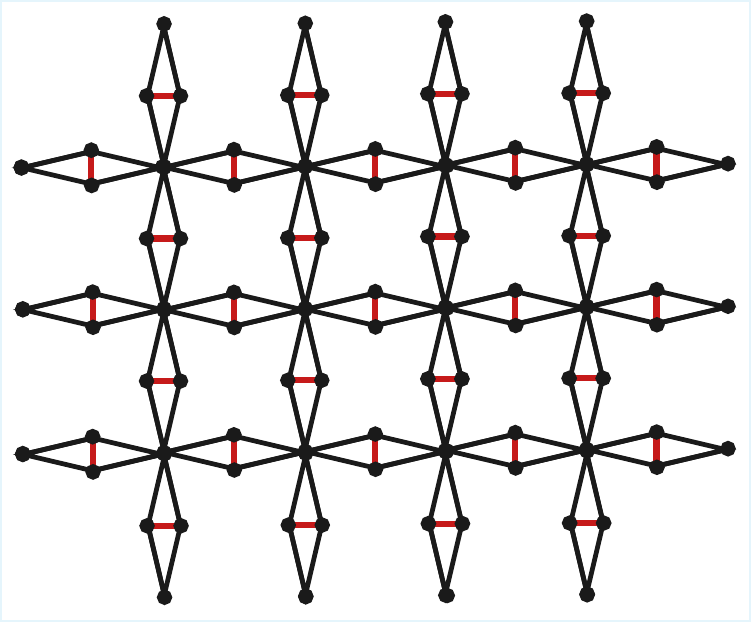}
\caption{Spin-$s$ Heisenberg model on the ideal diamond-decorated
square lattice. Black lines denote ferromagnetic exchange
interactions between monomer spins - $s$ and diagonal spins -
$\sigma$; red lines denote couplings between diagonal spins.}
\label{Fig_square_lattice}
\end{figure}

\begin{figure}[tbp]
\includegraphics[width=3in,angle=0]{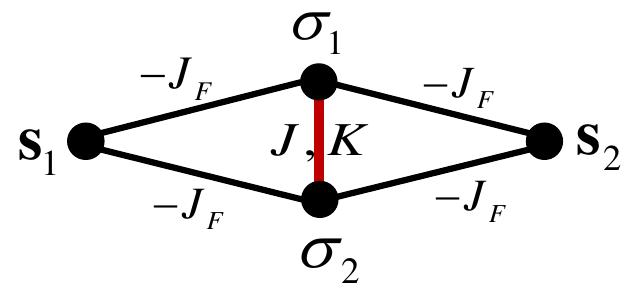}
\caption{Ideal diamond unit with monomer spins-$s$ and diagonal
spins-$\sigma$. Black lines denote ferromagnetic exchange
interactions $-J_F $ between monomer spins $\mathbf{s}_1,
\mathbf{s}_2$ and diagonal spins $\mathbf{\sigma}_{1},
\mathbf{\sigma}_{1}$. Red lines denote couplings between diagonal
spins, which includes bilinear and all relevant higher-order
terms.} \label{Fig_diamond}
\end{figure}

In this section, we investigate the ground state properties of the
spin-$s$ ferromagnetic-antiferromagnetic (F-AF) Heisenberg model
on several diamond-decorated lattices: the diamond chain, as well
as two- and three-dimensional lattices composed of ideal diamond
units. For concreteness, Fig.~\ref{Fig_square_lattice} illustrates
the diamond-decorated square lattice. The Hamiltonian can be
decomposed into a sum over local diamond units:
\begin{equation}
\hat{H}=\sum_{i}\hat{H}_{i},  \label{H}
\end{equation}%
where the sum runs over all $N_{b}=zN/2$ diamonds in the lattice
(with $N$ the number of lattice sites and $z$ the coordination
number). The local Hamiltonian $\hat{H}_{i}$ for the $i$-th
diamond, shown in Fig.~\ref{Fig_diamond}, reads
\begin{equation}
\hat{H}_{i}=-J_{F}(\mathbf{s}_{i}+\mathbf{s}_{i+1})\cdot (\mathbf{\sigma }%
_{1,i}+\mathbf{\sigma }_{2,i})+J\mathbf{\sigma }_{1,i}\cdot \mathbf{\sigma }%
_{2,i}+\sum_{n=2}^{2\sigma }K_{n}(\mathbf{\sigma }_{1,i}\cdot \mathbf{\sigma
}_{2,i})^{n}  \label{h}
\end{equation}%
where $\mathbf{s}_{i}$ is a spin-$s$ operator, and $\mathbf{\sigma
}_{1,i}$, $\mathbf{\sigma }_{2,i}$ are spin-$\sigma $ operators
associated with the $i$-th diamond. The interaction $J_{F}$
between the central monomer spins $\mathbf{s}_{i}$,
$\mathbf{s}_{i+1}$ and the diagonal spins $\mathbf{\sigma
}_{1,i}$, $\mathbf{\sigma }_{2,i}$ is ferromagnetic; we set
$J_{F}=1$ without loss of generality. In addition to the bilinear
antiferromagnetic coupling $J $ between the diagonal spins, we
include all relevant higher-order terms with couplings $K_{n}$.

A key observation from Eq.~(\ref{h}) is the local conservation of the
composite spin $\mathbf{L}_{i}=\mathbf{\sigma }_{1,i}+\mathbf{\sigma }_{2,i}$
on each diagonal. This composite spin is a conserved quantity with quantum
numbers $L=0,1,\dots ,2\sigma $. In terms of $\mathbf{L}_{i}$, the local
Hamiltonian can be rewritten as
\begin{equation}
\hat{H}_{i}=-(\mathbf{s}_{i}+\mathbf{s}_{i+1})\cdot \mathbf{L}_{i}+U(L_{i}),
\label{h1}
\end{equation}%
where
\begin{equation}
U(L)=J\left( \frac{1}{2}\mathbf{L}^{2}-\sigma(\sigma+1)\right) +\sum_{n=2}^{2\sigma
}K_{n}\left( \frac{1}{2}\mathbf{L}^{2}-\sigma(\sigma+1)\right) ^{n}  \label{UL}
\end{equation}%
is the internal energy of the composite spin, and $\mathbf{L}^{2}=L(L+1)$.

As follows from Eq.~(\ref{h1}), the model (\ref{H}) reduces to a
mixed-spin $(s,L)$ system on the corresponding Lieb lattice, with
ferromagnetic interactions between the $s$ and $L$ spins. However,
the composite spins $L_{i}$ can take different values in the range
$L_{i}=0,1,\dots ,2\sigma $ on different diamonds. Consequently,
the ground state is the fully polarized
ferromagnetic state with all spins $\mathbf{s}_{i}$ and
$\mathbf{L}_{i}$ aligned, except when $L_{i}=0$. In the case
$L_{i}=0$, the singlet on the diagonal of the $i$-th diamond is an
exact eigenstate independent of the states of the neighboring
monomer spins $\mathbf{s}_{i}$ and $\mathbf{s}_{i+1}$, which
effectively decouples the system at that location. The total
ground state energy is therefore
\begin{equation}
E=\sum_{i}\varepsilon _{i},  \label{EF}
\end{equation}%
with the energy of each diamond given by
\begin{equation}
\varepsilon _{i}=-2sL_{i}+U(L_{i}).  \label{Ei}
\end{equation}

Equation (\ref{EF}) implies that to determine the ground state
configuration of composite spins $\{L_{i}\}$, it suffices to
minimize each $\varepsilon _{i}$ individually. Since $\varepsilon
_{i}$ depends only on $L_{i}$ and not on neighboring composite
spins, all $\varepsilon _{i}$ can be minimized separately.
Moreover, because all $\varepsilon _{i}$ have the same functional
form, it is enough to consider a single diamond and select the
value $L_{gs}$ that minimizes Eq.~(\ref{Ei}). The ground state of
the total system is then given by the uniform configuration with
all composite spins $L_{i}=L_{gs}$, and the total energy is
$E=N_{b}\varepsilon _{gs}$, where
\begin{equation}
\varepsilon (L)=-2sL+U(L)  \label{eL}
\end{equation}%
is minimized over $L=0,1,\dots ,2\sigma $.

If the optimal composite spin is $L_{gs}=0$, the ground state wave
function of (\ref{H}) is a product of singlet states on all
diamond diagonals, while all monomer spins remain free, yielding a
ground state degeneracy of $(2s+1)^{N}$. When $L_{gs}\geq 1$, the
ground state is ferrimagnetic (or ferromagnetic for
$L_{gs}=2\sigma $) with total spin
$S_{\text{tot},gs}=Ns+N_{b}L_{gs}$, and is degenerate over
projections $S_{\text{tot}}^{z}=-S_{\text{tot},gs},\dots
,S_{\text{tot},gs}$. The wave function of (\ref{H}) with maximal
polarization $S_{\text{tot}}^{z}=S_{\text{tot},gs}$ is a direct
product of states where all monomer spins $\mathbf{s}_{i}$ point
up ($s_{i}^{z}=s$) and the diagonal spins $\mathbf{\sigma
}_{1,i}$, $\mathbf{\sigma }_{2,i}$ on each diamond form a state
with $(\mathbf{\sigma }_{1,i}+\mathbf{\sigma
}_{2,i})^{2}=L_{gs}\left( L_{gs}+1\right) $ and $\mathbf{\sigma
}_{1,i}^{z}+\mathbf{\sigma }_{2,i}^{z}=L_{gs}$.

As will be shown in Sec.~VI, the model exhibits a gapless
excitation spectrum in the sector
$S_{\text{tot}}<S_{\text{tot},gs}$ and a gap for excitations with
$S_{\text{tot}}>S_{\text{tot},gs}$, which require composite spins
$L>L_{gs}$. In this respect, the properties of the model resemble
those of conventional Lieb-Mattis ferrimagnets. The energy gap for
$S_{\text{tot}}>S_{\text{tot},gs}$ manifests in the magnetization
curves as plateaus and jumps, as will be shown in Sec.~V. Thus,
despite the apparent simplicity of the ground state, the model
possesses rich physical properties.

\subsection{Heisenberg model ($K_{n}=0$)}

We first consider the pure Heisenberg case, where all $K_{n}=0$.
In this simple case, the energy of a single diamond can be written
as
\begin{equation}
\varepsilon (L)=\frac{J}{2}\left( L-x\right) ^{2}-x^{2}-2J,
\end{equation}%
with
\begin{equation}
x=\frac{2s}{J}-\frac{1}{2}.
\end{equation}

The minimal energy $\varepsilon $ is achieved by the composite
spin value $L$ closest to $x$. Consequently, as the coupling $J$
increases, the parameter $x $ decreases, and the system undergoes
a cascade of $2\sigma $ transitions: it starts in the
ferromagnetic phase ($J<s/\sigma $) with $L=2\sigma $, then
successively passes through ferrimagnetic phases with $L=(2\sigma
-1)$, $2\sigma -2\ldots $, and so on, finally ending in the
monomer-dimer phase ($J>2s$) with $L=0$. Generally, the transition
between phases with $L$ and $L-1
$ occurs at%
\begin{equation}
J_{(L-1)/L}=\frac{2s}{L}.
\end{equation}

At all transition points except $J_{0/1}$, the ground state
degeneracy is $W\sim 2^{N_{b}}$, because each composite spin can
independently take either of the two values. At the point
$J_{0/1}$, the degeneracy is higher than $2^{N_{b}}$, since
singlet composite spins ($L=0$) effectively cut the system into
independent ferromagnetic clusters formed by spins $s$ and $L=1$,
separated by $L=0$ boundaries. In this case, the ground state
degeneracy for the diamond chain can be calculated analytically,
yielding $W\sim (2s+3)^{N}$ (see Sec.III for details), while for
higher-dimensional lattices it must be computed numerically.

\subsection{General case}

In the general case, the internal energy $U(L)$ is a polynomial in
$L$ of degree $2\sigma$. The ground state value of the composite
spin depends on $J$ and all $K_n$. The phase diagram in the
parameter space $(J, K_2, K_3, \dots, K_{2\sigma})$ is generally
complicated. We present the full phase diagram explicitly for the
case $s = \sigma = 1$ in the next subsection. However, we can make
one important general statement: in the parameter space, there
exists a special point where the energies of all composite spin
values $L = 0, 1, \dots, 2\sigma$ in Eq.(\ref{eL}) are equal. The position of this point is determined by the series of equalities
\begin{equation}
\varepsilon(L=0) = \varepsilon(L=1) = \dots = \varepsilon(L=2\sigma).
\label{eqs}
\end{equation}

These constitute $2\sigma $ linear equations in the $2\sigma $
parameters $(J,K_{2},\dots ,K_{2\sigma })$, which uniquely
determine a single special point where all conditions are
satisfied. At this point, the ground state degeneracy is maximal.
For the diamond chain, it can be calculated analytically (details
are in Sec.III), giving
\begin{equation}
W_{\max }\sim (2s+4\sigma +1)^{N}
\end{equation}%
while for higher-dimensional lattices numerical calculations are
required.

\subsection{Spin-$1$ model with bilinear and biquadratic interactions}

To make the general formalism more concrete, we now consider the
important special case $s=\sigma =1$, retaining only the bilinear
term $J$ and the biquadratic term $K_{2}\equiv K$ (all
higher-order couplings $K_{3},\ldots $ are irrelevant for $\sigma
=1$). This case is not only analytically tractable but also
relevant for real magnetic materials. The local Hamiltonian
reduces to
\begin{equation}
\hat{H}_{i}=-J_{F}(\mathbf{s}_{i}+\mathbf{s}_{i+1})\cdot (\mathbf{\sigma }%
_{1,i}+\mathbf{\sigma }_{2,i})+J\mathbf{\sigma }_{1,i}\cdot \mathbf{\sigma }%
_{2,i}+K(\mathbf{\sigma }_{1,i}\cdot \mathbf{\sigma }_{2,i})^{2}
\end{equation}

The composite spin $\mathbf{L}_{i}=\mathbf{\sigma
}_{1,i}+\mathbf{\sigma }_{2,i}$ now takes values $L=0,1,2$. The
internal energy becomes
\begin{equation}
U(L)=J\left( \frac{L(L+1)}{2}-2\right) +K\left( \frac{L(L+1)}{2}-2\right)
^{2}.
\end{equation}

Explicitly, the lowest diamond energies for $L = 0, 1, 2$ are:
\begin{align}
\varepsilon_0 &= -2J + 4K, \\
\varepsilon_1 &= -2 - J + K, \\
\varepsilon_2 &= -4 + J + K.
\end{align}

\begin{figure}[tbp]
\includegraphics[width=5in,angle=0]{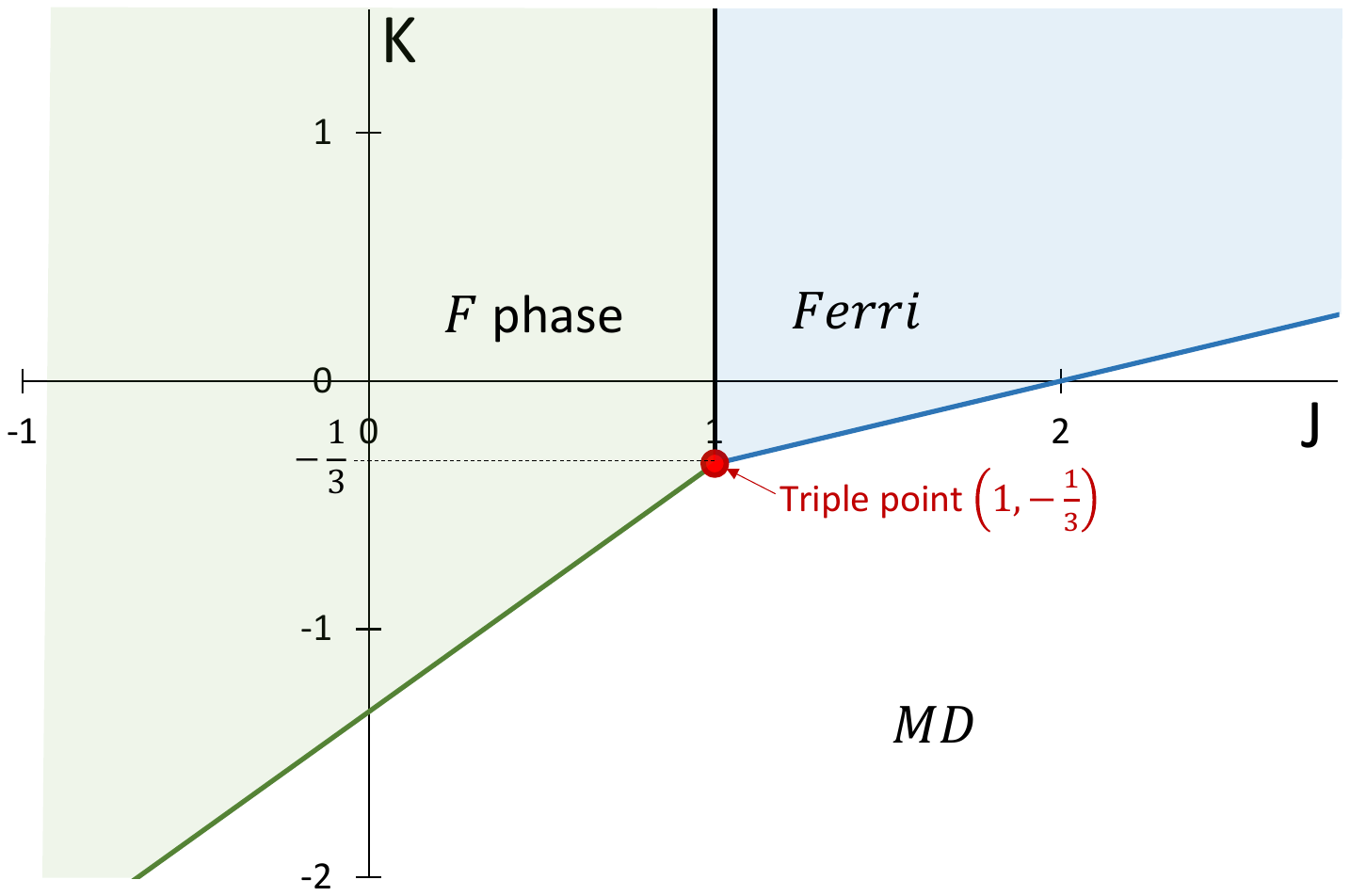}
\caption{Ground state phase diagram of the ideal diamond-decorated
spin-$1$ model in the $(J,K)$ plane. The three phases correspond
to composite spin $L=0$ (monomer-dimer), $L=1$ (ferrimagnetic),
and $L=2$ (ferromagnetic). Phase boundaries are determined by
equalities of the local ground state energies.}
\label{Fig_phase_diagram}
\end{figure}

By comparing these energies, we obtain the phase diagram shown in
Fig.~\ref{Fig_phase_diagram}. Three distinct phases emerge: the
monomer-dimer (MD) phase ($L=0$), the ferrimagnetic (Ferri) phase
($L=1$), and the ferromagnetic (F) phase ($L=2$). The phase
boundaries are given by:
\begin{align}
J &= 3K + 2 \quad (J > 1) & &\text{(MD/Ferri boundary)}, \\
J &= K + \frac{4}{3} \quad (J < 1) & &\text{(F/MD boundary)}, \\
J &= 1 \quad (K > -\tfrac{1}{3}) & &\text{(F/Ferri boundary)}.
\end{align}

All three phases meet at the triple point $(J,K)=(1,-1/3)$, where
$\varepsilon _{0}=\varepsilon _{1}=\varepsilon _{2}$. As shown in
Fig.~\ref{Fig_phase_diagram}, the critical line between the MD and
F phases, as well as the triple point, emerges only upon the
inclusion of the biquadratic interaction, highlighting its
essential role in stabilizing these features.

In contrast to the uniform ground states of the F-AF diamond
model, the antiferromagnetic spin-$1$ Heisenberg ideal diamond
chain (with $J_{F}=-1$) exhibits symmetry-broken ground state
phases in which the composite spins $L_{i}$ are not constant along
the chain \cite{Zoshki_2026, JPSJ.86.033707}. This difference in
ground state structure arises from the distinct spectra of the
local Hamiltonian $\hat{H}_{i}$ for the F-AF and purely AF cases.

\section{Ground state degeneracy of the ideal diamond chain}

The ground state of the Hamiltonian (\ref{H}) is degenerate within
each bulk phase. In the MD phase, where every diagonal is in the
singlet state $L_i = 0 $, the $N$ monomer spins remain free. This
yields a macroscopic ground state degeneracy $W_{\mathrm{MD}} =
(2s+1)^N$, corresponding to the independent spin states of each
free monomer. In a ferrimagnetic phase characterized by a fixed
composite spin $L$, the total spin of the system is
$S_{\text{tot}} = N s + N_b L$, which gives a multiplet degeneracy
$W_L = (z L + 2s)N + 1$. In the fully polarized ferromagnetic
phase, all diamonds are in the maximal composite spin state $L_i =
2\sigma$, resulting in $S_{\text{tot}} = N s + z \sigma N$ and a
degeneracy $W_{\mathrm{F}} = 2N(z\sigma + s) + 1$. Thus,
macroscopic ground state degeneracy in the bulk phases occurs only
in the MD phase.

Of particular interest is the ground state degeneracy on the phase
boundaries, where the system becomes macroscopically degenerate.
As established in Ref.~\cite{KDNDR}, such macroscopic degeneracy
arises when the local Hamiltonian possesses multiple degenerate
ground states - precisely the situation along the critical lines
of the present model.

The counting of ground states on phase boundaries was previously
addressed for the spin-$\frac{1}{2}$ Heisenberg model on ideal
diamond-decorated lattices in Refs.~\cite{diamond1d, DKV2D}. For
the diamond chain, this counting can be performed exactly, while
for higher-dimensional lattices it reduces to a percolation
problem. We now apply these methods to the spin-$s$ model under
consideration.

\subsection{Boundary between MD and ferrimagnetic phases}

We begin with the diamond chain. As a representative example,
consider the critical line separating the MD and ferrimagnetic
phases, where singlets ($L_{i}=0$) and triplets ($L_{i}=1$)
coexist. In this case, singlets act as effective separators,
partitioning the chain into finite ferromagnetic clusters of
consecutive monomer spins and diagonal triplets. Any ground state
configuration can thus be viewed as a sequence of randomly
distributed singlet diamonds and intervening ferromagnetic
clusters.

Following the approach of Ref.~\cite{diamond1d}, the total number
of ground states $W_{N}$ for an open chain of $N$ diamonds
satisfies the recurrence relation
\begin{equation}
W_{N}=\sum_{k=1}^{N-1}f(k)W_{N-k}+(2s+1)f(N)+f(N+1),  \label{rec}
\end{equation}%
where $f(k)$ denotes the number of multiplets contributed by a
cluster of $(k-1)$ diamonds (i.e., a block of $k-1$ consecutive
monomers and triplets bounded by singlets or chain ends), the term
$(2s+1)f(N)$ describes the configuration with one singlet on the
last diamond, and the term $f(N+1)$ accounts for the configuration
with no singlets at all.

For a ferromagnetic cluster of $(k-1)$ diamonds with $L=1$, the
maximal spin is $S_{\text{max}}=ks+k-1$, and one finds
$f(k)=2(s+1)k-1$. Solving Eq.~(\ref{rec}) with this expression
yields
\begin{equation}
W_{0/1}=4(s+1)^{2}(2s+3)^{N-1}.
\end{equation}

For the special case $s = \sigma = 1$, this reduces to
\begin{equation}
W_{\mathrm{MD/Ferri}} = \frac{16}{5} \cdot 5^N.  \label{L01}
\end{equation}
Remarkably, this result holds along the entire MD/Ferri phase boundary,
including the point $(J=2, K=0)$.

\subsection{Boundary between F and MD phases}

For the special case $s = \sigma = 1$ on the critical line between
the F and MD phases, the relevant local states are singlets ($L_i
= 0$) and quintets ($L_i = 2$). A ferromagnetic cluster of $(k-1)$
consecutive quintets (and monomer spins between them) has $f(k) =
6k - 3$ multiplets. Substituting into Eq.~(\ref{rec}) gives the
ground state degeneracy
\begin{equation}
W_{\mathrm{MD/F}} = a \cdot b^N,  \label{L02}
\end{equation}
with $b = \frac{\sqrt{33}+5}{2} \approx 5.372$ and $a \approx 3.255$.

\subsection{Multicritical point: maximal degeneracy}

The largest degeneracy occurs at the multicritical point where all
composite spin states are degenerate: $\varepsilon_0 =
\varepsilon_1 = \dots = \varepsilon_{2\sigma}$. Here, singlets
separate clusters that may contain arbitrary composite spins $L_i
= 1, 2, \dots, 2\sigma$ in any order. For a ferromagnetic cluster
of $(k-1)$ diamonds, the total spin depends on the configuration
$\{L_i\}$ as
\begin{equation}
S_{\text{tot}} = ks + \sum_{i=1}^{k-1} L_i.
\end{equation}
The degeneracy for a particular configuration $\{L_i\}$ is $w =
2S_{\text{tot}} + 1$. Summing over all configurations gives
\begin{equation}
f(k) = \sum_{L_1=1}^{2\sigma} \sum_{L_2=1}^{2\sigma} \cdots
\sum_{L_{k-1}=1}^{2\sigma} \left( 2ks + 1 + 2\sum_{i=1}^{k-1} L_i \right).
\end{equation}

Performing the sums iteratively, we obtain
\begin{equation}
f(k)=(2\sigma )^{k-1}\left[ 2ks+1+(k-1)(2\sigma +1)\right] .
\end{equation}

Solving Eq.(\ref{rec}) with this $f(k)$ yields
\begin{equation}
W_{\max }=(2s+2\sigma +1)^{2}(2s+4\sigma +1)^{N-1}.  \label{Wmax}
\end{equation}

For the special case $s = \sigma = 1$, this corresponds to the
triple point and reduces to
\begin{equation}
W_{\mathrm{TP}} = \frac{25}{7} \cdot 7^N.  \label{L012}
\end{equation}

As shown in Ref.~\cite{diamond1d}, for cyclic chains with $s = \sigma = \frac{1}{2}$ the exponent
in $W$ remains the same, while the prefactor changes and
additional exponentially small terms may appear. However, these
changes are irrelevant for the residual entropy in the
thermodynamic limit, so the results obtained for open chains
provide the correct asymptotic behavior.

In Ref.~\cite{diamond1d} it was also shown that the ground state
degeneracy corresponds to that of a system of independent
spin-$\frac{3}{2}$ degrees of freedom, analogous to the behavior
observed in the special version of distorted diamond chain. As
follows from Eq.~(\ref{Wmax}), this nontrivial property
generalizes to arbitrary spins $s$ and $\sigma$, with the ground
state degeneracy matching that of independent spins-$(s +
2\sigma)$. However, as demonstrated for $s = \sigma = \frac{1}{2}$
in Ref.~\cite{DKV2D}, this remarkable property does not survive on
two- and three-dimensional diamond-decorated lattices. As will be
shown in Sec.~IV for the general case of arbitrary $s$ and
$\sigma$, this property likewise fails in higher dimensions.

\subsection{Boundary between ferrimagnetic phases}

Finally, on the boundary between phases with composite spins $L$
and $L+1$, the degeneracy for a cyclic chain can be calculated
directly as
\begin{align}
W_{L/(L+1)}& =\sum_{k=0}^{N}\binom{N}{k}\left[ 2N(s+L)+2k+1\right] \\
& =\left[ 2N(s+L)+N+1\right] 2^{N}.
\end{align}

Thus, for the special case $s = \sigma = 1$ on the critical line
between the F and ferrimagnetic phases (including the point $(J=1,
K=0)$), the ground state degeneracy is
\begin{equation}
W_{\mathrm{F/Ferri}} = (5N + 1) 2^N.  \label{L12}
\end{equation}

Remarkably, this expression admits a straightforward
generalization to any lattice with coordination number $z$:
\begin{equation}
W_{\mathrm{F/Ferri}}=(3N_{b}+2N+1)2^{N_{b}},  \label{W_F/Ferri}
\end{equation}%
where $N_{b}=zN/2$ is the number of diamonds.

\section{Ground state degeneracy on diamond-decorated lattices}

In this section, we focus on the special case $s=\sigma =1$ with
bilinear and biquadratic interactions, which serves as a concrete
illustration of the general formalism developed in Section II.
While the preceding analysis applies to arbitrary spin values $s$
and $\sigma$, the explicit calculation of ground state
degeneracies on two- and three-dimensional lattices becomes
technically involved for general spins. Nevertheless, the key
physical insight - namely, the mapping of the degeneracy counting
problem onto a bond percolation framework - remains valid for any
$s$ and $\sigma$. The percolation approach presented here for
$s=\sigma =1$ can be straightforwardly generalized to higher spins
by appropriately modifying the cluster weight functions, which
depend on the spin values through the possible composite spin
states $L=0,1,\dots ,2\sigma $ and their associated degeneracies.
Thus, while we present numerical results for the spin-1 case, the
methodology is fully general and can be applied to study other
spin combinations.

The recurrence approach used for the diamond chain cannot be
directly extended to two- and three-dimensional diamond-decorated
lattices. In higher dimensions, the problem of counting ground
state degeneracies on the critical lines maps naturally onto a
bond percolation problem. Our analysis follows the methodology
developed in Ref.~\cite{DKV2D} for the spin-$\frac{1}{2}$
Heisenberg model on ideal diamond-decorated lattices; here we
briefly outline the key steps adapted to the spin-$1$ case.

As established earlier, a singlet on a diamond diagonal ($L_i =
0$) effectively decouples the two neighboring monomer spins,
thereby cutting the bond between them. Consequently, each exact
ground state of the total Hamiltonian $\hat{H}$ is fully
characterized by a configuration of diamonds where some diagonals
form singlets and others form either triplets ($L_i = 1$) or
quintets ($L_i = 2$), depending on the critical line under
consideration. To compute the total ground state degeneracy, one
must sum the degeneracy contributions from all possible singlet
configurations.

This problem is isomorphic to bond percolation: diamonds with
triplet or quintet diagonals correspond to connected bonds, while
diamonds with singlet diagonals correspond to disconnected (or
"broken") bonds. For a given configuration $\omega _{K}$
containing $K$ connected bonds (i.e., $K$ diamonds in a nonsinglet
state), the lattice decomposes into disconnected ferromagnetic
clusters. Each cluster consists of a set of monomer spins linked
by intervening triplet or quintet diamonds. Consider the $i$-th
such cluster, containing $n_{i}$ monomer spins and $l_{i}$
connected bonds. The total number of spins in the cluster is
$n_{i}+2l_{i}$, and its ground state degeneracy depends on the
specific phase boundary under study.

On the critical line between the ferromagnetic (F) and
monomer-dimer (MD) phases, each cluster is fully polarized, with
total spin $S_{\text{tot},i}=n_{i}+2l_{i}$. The corresponding
ground state degeneracy, given by the number of multiplets, is
$2n_{i}+4l_{i}+1$. On the critical line between the ferrimagnetic
(Ferri) and MD phases, each cluster has total spin
$S_{\text{tot},i}=n_{i}+l_{i}$, yielding a degeneracy
$2n_{i}+2l_{i}+1$. At the triple point, where all three phases
meet, each connected bond within a cluster can independently be
either a triplet or a quintet. Summing over all $2^{l_{i}}$
configurations of triplets or quintets within the cluster gives a
cluster degeneracy $(2n_{i}+3l_{i}+1)2^{l_{i}}$.

For a given configuration $\omega _{K}$, the total number of
ground states is the product of the degeneracies of all its
constituent clusters:
\begin{eqnarray}
W_{\mathrm{MD/F}}(\omega _{K},N) &=&\prod_{i\in \omega _{K}}(2n_{i}+4l_{i}+1)
\label{W_MD/F} \\
W_{\mathrm{MD/Ferri}}(\omega _{K},N) &=&\prod_{i\in \omega
_{K}}(2n_{i}+2l_{i}+1)  \label{W_MD/Ferri} \\
W_{\mathrm{TP}}(\omega _{K},N) &=&\prod_{i\in \omega
_{K}}(2n_{i}+3l_{i}+1)2^{l_{i}}  \label{W_TP}
\end{eqnarray}

To obtain the total degeneracy for a given number $K$ of connected
bonds, we sum over all geometrically distinct configurations
$\omega _{K}$ on an $N$-site lattice:
\begin{equation}
W_{\alpha }(K,N)=\sum_{\omega _{K}}W_{\alpha }(\omega _{K},N),  \label{ZK}
\end{equation}%
where $\alpha $ labels the phase boundary (MD/F, MD/Ferri) or the
triple point (TP). Finally, the full ground state degeneracy is
obtained by summing over all possible $K$, i.e., over all numbers
of connected bonds from $0$ to the total number of diamonds
$N_{b}$:
\begin{equation}
W_{\alpha }(N)=\sum_{K=0}^{N_{b}}W_{\alpha }(K,N)  \label{W}
\end{equation}

The evaluation of these sums for specific two- and
three-dimensional lattices reduces to solving a bond percolation
problem, with the cluster weights given by the expressions above.
This approach provides a systematic way to compute the macroscopic
ground state degeneracies along the critical lines and at the
triple point of the spin-$1$ diamond-decorated model.

\begin{figure}[tbp]
\includegraphics[width=5in,angle=0]{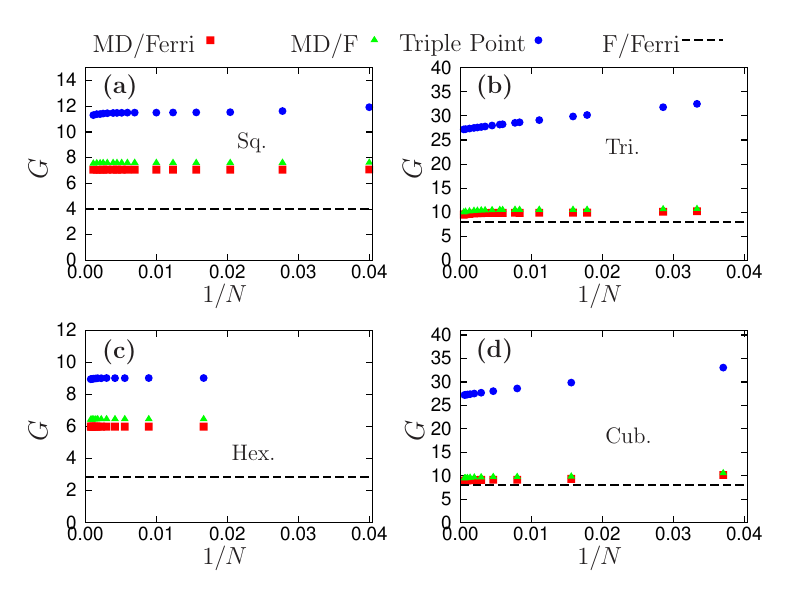}
\caption{The values of $G=W^{1/N}$ as a function of $1/N$ for (a)
square; (b) triangular; (c) honeycomb; and (d) cubic lattices on
phase boundaries MD/Ferri, MD/F, F/Ferri and in the triple point.}
\label{fig:zn}
\end{figure}

Using the numerical Monte-Carlo approach details in
Ref.\cite{DKV2D}, we find that for all studied lattices
(hexagonal, square, triangular, cubic), the ground state
degeneracy grows exponentially with $N$
\begin{equation}
W_{\alpha }(N)\sim G_{\alpha }^{N}
\end{equation}%
where the base $G_{\alpha }$ is a lattice-dependent exponent
characterizing the asymptotic growth. The finite-size behavior of
$G_{\alpha }(N)$ as a function of $1/N$ for various lattices is
shown in Figs.~\ref{fig:zn}.

The exponential growth of the ground states degeneracy gives rise
to a residual entropy per spin, defined as
\begin{equation}
\mathcal{S}_{0,\alpha }(N)=\frac{\ln (W_{\alpha }(N))}{\mathcal{N}}
\label{res_entropy}
\end{equation}%
with $\mathcal{N}=N+zN$ denoting the total number of spins-$1$. In
the thermodynamic limit $\mathcal{N}\to \infty $, the residual
entropy converges to a finite value:
\begin{equation}
\mathcal{S}_{0,\alpha }=\frac{\ln G_{\alpha }}{z+1}
\end{equation}

Table 1 lists these limiting values of $\mathcal{S}_{0}$, for
diamond-decorated models on hexagonal, square, triangular and
cubic lattices for three phase boundaries and for triple point.

\begin{table}[tbp]
\caption{Residual entropy per spin, $\mathcal{S}_{0,\alpha }$, for
ideal diamond-decorated lattices, on the phase boundaries MD/F,
MD/Ferri,
F/Ferri, and at the triple point}%
\begin{ruledtabular}
\begin{tabular}{ccccc}
Lattice & MD/F  & MD/Ferri & F/Ferri  & Triple point  \\
\hline Chain ($z=2$) & $0.560$ & $0.536$ & $0.231$ & $0.649$ \\
Hexagonal ($z=3$)    & $0.448(1)$ & $0.466(1)$ & $0.260$ & $0.550(1)$ \\
Square ($z=4$)       & $0.391(1)$ & $0.405(1)$ & $0.277$ & $0.488(1)$ \\
Triangular ($z=6$)   & $0.326(1)$ & $0.336(1)$ & $0.297$ & $0.473(1)$ \\
Cubic ($z=6$)        & $0.315(1)$ & $0.324(1)$ & $0.297$ & $0.472(1)$ \\
\end{tabular}
\end{ruledtabular}
\end{table}

Several notable trends emerge from Table 1. First, the residual
entropy at the triple point consistently exceeds that on any of
the phase boundaries for all lattices considered, highlighting the
enhanced ground state degeneracy where all three local states
coexist. Second, the coordination number $z$ exhibits contrasting
effects on different phase boundaries: while the residual entropy
generally decreases with increasing $z$ for the MD/F and MD/Ferri
boundaries and at the triple point, it shows a modest increase
with $z$ along the F/Ferri boundary, consistent with the
analytical expression in Eq.~(\ref{W_F/Ferri}). Third, the
observed residual entropies are remarkably high, particularly at
the triple point where they reach approximately 40--60\% of the
maximal possible entropy $\ln 3$ corresponding to completely free
$s=1$ spins.

\section{Ground state magnetization}

We now investigate the ground state magnetization of the model
described by Hamiltonian (\ref{H}) for the case $s=\sigma =1$ at
zero temperature. Within the three bulk phases, the magnetization
per spin takes simple characteristic values. In the ferromagnetic
(F) phase, all spins are fully aligned, yielding a saturation
magnetization per spin $m_{\mathrm{F}}=1$. In the monomer-dimer
(MD) phase, the system consists of independent monomer spins
decoupled by singlet dimers on the diagonals; since each singlet
carries zero magnetic moment, the overall magnetization vanishes,
$m_{\mathrm{MD}}=0$. In the ferrimagnetic (Ferri) phase, the
ground state total spin is $S_{\text{tot}}=N_{b}+N$, leading to a
magnetization per spin $m_{\mathrm{Ferri}}=\frac{z+2}{2z+2}$,
which interpolates between the ferromagnetic and singlet limits as
a function of the coordination number $z$.

Of greater interest is the behavior of zero temperature
magnetization on the critical lines, where the ground state
manifold becomes macroscopically degenerate. To characterize the
magnetic response in such regimes, we define the total
magnetization $M$ as an average over all degenerate ground states:
\begin{equation}
M^{2}=\frac{1}{W}\sum_{i=1}^{W}r(i)  \label{M}
\end{equation}%
where $r(i)=\left\langle \psi _{i}\right\vert
\mathbf{S}_{tot}^{2}\left\vert \psi _{i}\right\rangle $,
$\mathbf{S}_{tot}=\sum \mathbf{s}_{j}$ is the total spin operator
of the system, and the sum runs over all $W$ ground states $|\psi
_{i}\rangle $. For infinite lattices, this definition effectively
captures the long-range spin correlations $\left\langle
\mathbf{s}_{\mathbf{i}}\cdot \mathbf{s}_{\mathbf{j}}\right\rangle
$ as $|\mathbf{i}-\mathbf{j}|\rightarrow \infty $, averaged
uniformly over the entire ground state manifold.

On the critical line separating the ferromagnetic and
ferrimagnetic phases, the magnetization can be calculated
analytically for any lattice. In this case, the ground state
manifold comprises ferromagnetic states corresponding to all
possible configurations in which each diamond diagonal is
independently in either a triplet ($L_{i}=1$) or quintet
($L_{i}=2$) state. For a configuration with $k$ triplets and
$N_{b}-k$ quintets, the total spin is $S_{\text{tot}}=(z+1)N-k$,
and each such multiplet contributes a degeneracy factor
$2S_{\text{tot}}+1$. The corresponding value of $r(k)$ for this
configuration is therefore
$r(k)=S_{\text{tot}}(S_{\text{tot}}+1)(2S_{\text{tot}}+1)$.

Summing over all configurations, the total numerator in Eq.~(\ref{M})
becomes
\begin{equation}
R(N)=\sum_{k=0}^{N_{b}}C_{N_{b}}^{k}r(k)
\end{equation}

In the thermodynamic limit $N\gg 1$, one finds the asymptotic
behavior $R(N)=2^{N_{b}+1}(\frac{3z}{4}+1)^{3}N^{3}$, while the
total number of ground states scales as
$W(N)=2^{N_{b}}(\frac{3z}{2}+2)N$. Combining these results yields
the magnetization
\begin{equation}
M_{\mathrm{F/Ferri}}=\left( \frac{3z}{4}+1\right) N  \label{M_F_Ferri}
\end{equation}%
which is precisely the average of the magnetizations of the pure
ferromagnetic and ferrimagnetic phases. This linear dependence on
the system size reflects the extensive nature of the total spin on
this critical line and demonstrates that the ground state manifold
supports a finite magnetization per spin even in the presence of
macroscopic degeneracy.

The result (\ref{M_F_Ferri}) is valid for any lattice on the
critical line between the ferromagnetic and ferrimagnetic phases.
For the other phase boundaries, however, the magnetization
exhibits a nontrivial dependence on the lattice geometry. We first
consider the diamond chain.

On the critical line between the singlet (MD) and ferrimagnetic
phases, the ground state consists of a random distribution of
singlet ($L_{i}=0$) and triplet ($L_{i}=1$) diamonds. Singlets act
as effective separators, partitioning the chain into independent
ferromagnetic clusters composed of consecutive diamonds. Any
ground state configuration can therefore be represented as a
sequence of randomly distributed singlets and intervening clusters
with triplet diagonals. The total numerator $R(N)$ is then
obtained by summing over contributions from individual clusters.
Following an approach analogous to the recurrence relation
(\ref{rec}), the function $R(N)$ satisfies the following equation
\begin{equation}
R(N)=\sum_{k=1}^{N}[g(k)W_{N-k}+R(N-k)]f(k)+g(N+1)  \label{RN}
\end{equation}%
where $g(k)=S(k)(S(k)+1)$, $S(k)$ is the total spin of a cluster,
and $f(k)$ is the number of multiplets for a cluster of $(k-1)$
diamonds as defined previously. On this critical line, $S(k)=2k-1$
and $g(k)=(2k-1)2k$. Solving Eq.~(\ref{RN}) in the limit $N\gg 1$
yields $R(N)\sim NW_{N}$, with $W_{N}\sim 5^{N}$. Consequently,
$M\sim \sqrt{N}$ as $N\to \infty $, implying that the
magnetization per spin vanishes on the MD/Ferri critical line.

A similar analysis for the critical line between the ferromagnetic
and MD phases, as well as for the triple point, shows that $R(N)
\sim N W_N$ in all cases. Hence, for the ideal diamond chain, the
magnetization vanishes on all critical lines except the F/Ferri
boundary, where it remains finite. Thus, the magnetization
exhibits a discontinuity across all critical lines.

Extending the analysis to two- and three-dimensional
diamond-decorated lattices, the problem can again be mapped onto a
bond percolation framework, analogous to the treatment of ground
state degeneracy. Consider a particular configuration $\omega
_{K}$ with $K$ connected bonds (i.e., diamonds in nonsinglet
states). The lattice decomposes into independent clusters; the
$i$-th cluster contains $n_{i}$ monomer sites and $l_{i}$
connected bonds, and carries total spin $S_{i}$. The total number
of ground states for this configuration is given by Eq.~(\ref{W}),
and importantly, the expectation value $\mathbf{S}_{tot}^{2}$ is
the same for all ground states within that configuration. Because
different clusters are independent, we have
\begin{equation}
\left\langle \psi _{k}\right\vert \mathbf{S}_{tot}^{2}\left\vert \psi
_{k}\right\rangle =\left\langle \psi _{k}\right\vert \sum_{i\in \omega _{K}}%
\mathbf{S}_{i}^{2}\left\vert \psi _{k}\right\rangle  \label{S2}
\end{equation}%
since cross terms $\left\langle \psi _{k}\right\vert \mathbf{S}_{i}\cdot
\mathbf{S}_{j}\left\vert \psi _{k}\right\rangle =0$ for $i\neq j$. The
magnetization per spin for a given configuration $\omega _{K}$ is then
\begin{equation}
m^{2}(N,\omega _{K})=\frac{1}{\mathcal{N}^{2}}\sum_{i\in \omega
_{K}}S_{i}(S_{i}+1)  \label{mK}
\end{equation}

Averaging over the entire ground state manifold yields the overall
magnetization per spin:
\begin{equation}
m^{2}(N)=\frac{\sum_{K=0}^{N_{b}}\sum_{\omega _{K}}W(N,\omega
_{K})m^{2}(N,\omega _{K})}{\sum_{K=0}^{N_{b}}\sum_{\omega _{K}}W(N,\omega
_{K})}  \label{m}
\end{equation}

The behavior of $m(N)$ is governed by the cluster size
distribution. If a configuration $\omega_K$ contains only finite
clusters whose number scales with $N$, each cluster has total spin
of order unity, and the magnetization per spin vanishes as $m \sim
N^{-1/2}$. Conversely, if a configuration contains an infinite
percolation cluster encompassing a finite fraction $P$ of all
spins, then $m \sim P$ in the thermodynamic limit.

\begin{figure}[tbp]
\includegraphics[width=5in,angle=0]{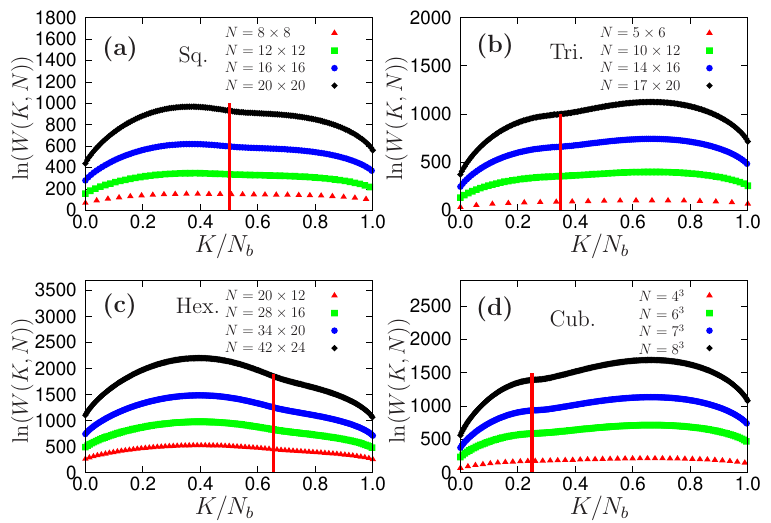}
\caption{Contributions to the partition function $\ln W(K,N)$
(Eq.~(\ref{ZK})) at the triple point as a function of the
connected-bond fraction $p = K/N_b$ for different system sizes $N$
and lattices: (a) square; (b) triangular; (c) hexagonal; (d)
cubic. Vertical red lines indicate the corresponding
bond-percolation thresholds, $p_c$.} \label{fig:ZB}
\end{figure}

According to percolation theory, an infinite cluster exists only
above the bond-percolation threshold $p_{c}$, which depends on the
lattice geometry. For $p>p_{c}$, the infinite cluster fraction
rapidly grows as $P\sim (p-p_{c})^{\beta }$, with $\beta \simeq
0.14$ in two dimensions and $\beta \simeq 0.4$ in three dimensions
\cite{percolation_review,beta04_percolation}. Therefore, the
presence of finite magnetization in the thermodynamic limit hinges
on whether the peak position $p_{0}=K_{\text{max}}/N_{b}$ of the
weight function $W(K,N)$ lies above $p_{c}$.

Our calculations reveal that for the MD/F and MD/Ferri boundaries,
$p_{0}$ lies substantially below $p_{c}$ for all studied lattices.
Consequently, the magnetization per spin vanishes as $m(N)\sim
N^{-1/2}$. At the triple point, however, the situation is
lattice-dependent. As shown in Fig.~\ref{fig:ZB}, the weight
function $W(K,N)$ peaks at $p_{0}$ well below $p_{c}$ for
hexagonal ($p_c=0.653$) and square lattices ($0.5$), indicating that configurations
containing an infinite cluster are exponentially suppressed.
Hence, the ground state magnetization vanishes for these lattices.
In contrast, for triangular and cubic lattices, $p_{0}\simeq 0.7$
is significantly higher than the respective percolation thresholds
$p_{c}\simeq 0.347$ for triangular \cite{tri_hex_percolation} and
$p_{c}\simeq 0.249$ for cubic \cite{cubic_percolation} lattice. In
these cases, the vast majority of ground states contain an
infinite cluster, leading to a finite magnetization in the
thermodynamic limit.

\begin{figure}[tbp]
\includegraphics[width=5in,angle=0]{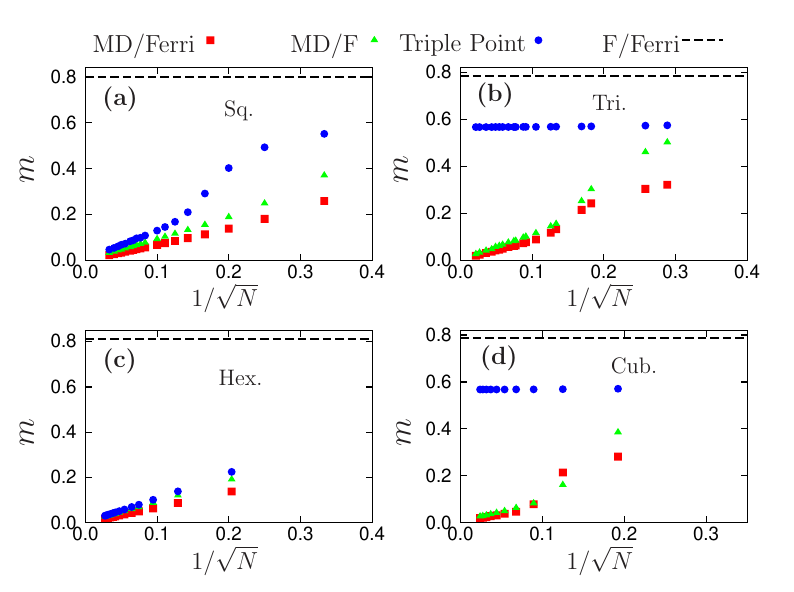}
\caption{Magnetization per spin calculated according to
Eq.~(\ref{m}) as a function of $N^{-1/2}$ for (a) square, (b)
triangular, (c) honeycomb, and (d) cubic lattices on the phase
boundaries MD/Ferri, MD/F, F/Ferri, and at the triple point.}
\label{fig:M}
\end{figure}

These predictions are confirmed numerically in Fig.~\ref{fig:M}.
For all cases except the triple point on triangular and cubic
lattices, the magnetization per spin scales as $m(N)\sim
N^{-1/2}$, extrapolating to zero in the thermodynamic limit. At
the triple point on triangular and cubic lattices, however, the
magnetization approaches a finite value $m\simeq 0.57$ as
$N\rightarrow \infty $. Remarkably, this value lies very close to
the magnetization of the ferrimagnetic phase, which for triangular
and cubic lattices ($z=6$) is
$m_{\mathrm{Ferri}}=\frac{4}{7}\simeq 0.571$. For the F/Ferri
boundary, the exact result (\ref{M_F_Ferri}) gives
$m_{\mathrm{F/Ferri}}=\frac{1}{2}m_{\mathrm{F}}+\frac{1}{2}m_{\mathrm{Ferri}}$, shown
by black dashed lines in Fig.~\ref{fig:M}.

\subsection{Low-temperature magnetization curves}

We now examine the full magnetization curve in an external
magnetic field $h$ at low temperature. The behavior depends
sensitively on both the phase and the dimensionality of the
lattice.

\subsubsection{Ferromagnetic phase}

In the ferromagnetic phase, the magnetization response is governed
by dimensionality. For three-dimensional lattices, robust
long-range ferromagnetic order persists, and an infinitesimal
field suffices to achieve saturation magnetization
$m_{\mathrm{F}}=1$. In two dimensions, however, the Mermin-Wagner
theorem precludes long-range order at finite temperature.
Consequently, the magnetization increases rapidly as $m=\chi h$
with high susceptibility $\chi \sim \exp ($const.$/T)$, before
eventually saturating at $m_{\mathrm{F}}=1$.

\subsubsection{Ferrimagnetic phase}

At zero temperature, the polarized monomer spins together with the
diamond triplets yield a magnetization
$m_{\mathrm{Ferri}}=\frac{z+2}{2z+2}$. A finite energy gap
separates the ground state from excited states, associated with
the conversion of a composite spin from $L=1$ to $L=2$. As a
result, the magnetization exhibits a plateau at
$m_{\mathrm{Ferri}}$ for fields $0\leq h\leq h_{1,2}$, where
$h_{1,2}=2(J-1)$. At $h=h_{1,2}$, all composite spins undergo a
transition from $L=1$ to $L=2$, leading to a discontinuous jump
from $m_{\mathrm{Ferri}}$ to saturation $m_{\mathrm{F}}=1$.

As in the ferromagnetic case, the initial magnetization
$m_{\mathrm{Ferri}}$ remains stable at low temperatures for
three-dimensional lattices, while for two-dimensional lattices it
exhibits field-dependent behavior $m \sim h
\exp(\text{const.}/T)$. The magnetization jump at $h = h_{1,2}$ is
smoothed over a temperature-dependent width $\Delta h \sim T$.

\subsubsection{Monomer-dimer (MD) phase}

At zero temperature, the free monomer spins respond immediately to
an external field, producing an initial magnetization
$m_{\mathrm{MD}} = \frac{1}{z+1}$. An energy gap above the ground
state manifold arises from the conversion of a composite spin from
$L = 0$ to $L = 2$ (for $J < 1$). The behavior depends on the
value of $J$:

\noindent \textbf{Case $J<1$: }Magnetization has a plateau at
$m=m_{\mathrm{MD}}$ for fields $0\leq h\leq h_{0,2}$, where
$h_{0,2}=3J-3K-4$. For $h>h_{0,2}$, the system saturates at $m=1$.

\noindent \textbf{Case $J>1$:} The transition from $L=0$ to $L=2$
proceeds via an intermediate $L=1$ state, resulting in two
successive magnetization jumps. The first jump is from plateau at
$m=m_{\mathrm{MD}}$ to plateau at $m=m_{\mathrm{Ferri}}$ at
$h=h_{0,1}=J-3K-2$, where all composite spins convert from $L=0$
to $L=1$. The second jump happens from plateau at
$m=m_{\mathrm{Ferri}}$ to $m=1$ at $h=h_{1,2}$ as before.

At low but nonzero temperatures, the free monomer spins produce a
linear paramagnetic response at weak fields, $m \sim h/T$. Once
the magnetization reaches $m_{\mathrm{MD}}$, it exhibits a plateau
that persists up to either $h_{0,2}$ (for $J < 1$) or $h_{0,1}$
(for $J > 1$). The jumps at these critical fields are broadened by
temperature over a scale $\Delta h \sim T$.

\begin{figure}[tbp]
\includegraphics[width=5in,angle=0]{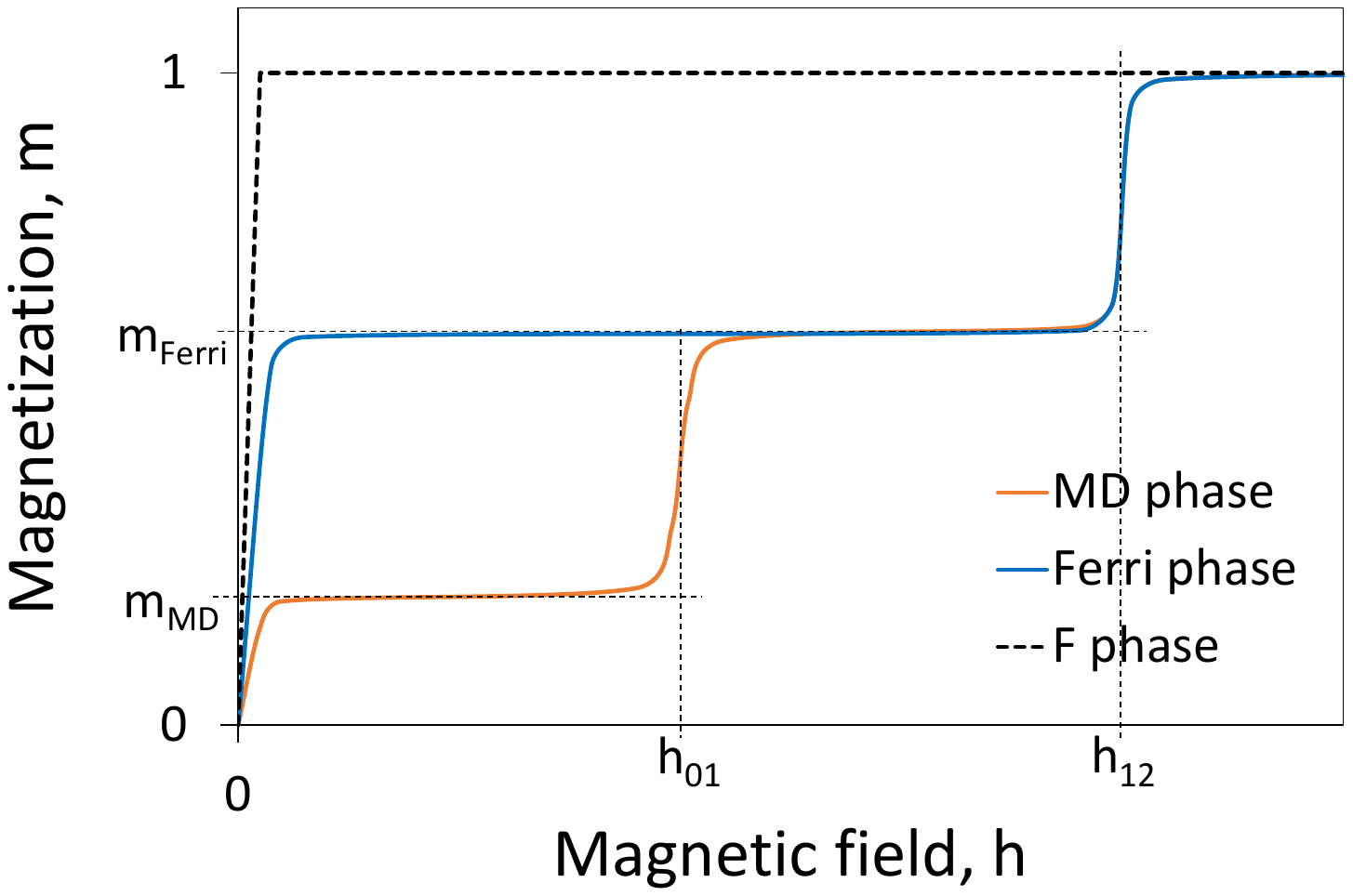}
\caption{Schematic magnetization curves in the low-temperature
limit for the ferromagnetic phase (black long-dashed line), the
ferrimagnetic phase (blue line), the MD phase (red dashed line).}
\label{fig:M_h}
\end{figure}

The magnetization curves in three phases are summarized and
schematically shown in Fig.\ref{fig:M_h}.

\subsubsection{Behavior on critical lines}

We now turn to the magnetization behavior on the critical lines.

\paragraph{F/Ferri phase boundary.}

On this boundary, the system is characterized by a ferromagnetic
state comprising central spins and a random distribution of
composite spins in $L = 1$ and $L = 2$ states. The truncated
partition function constructed from the ground state manifold is
$Z_0 = e^{Nh/T}(e^{h/T} + e^{2h/T})^{N_b}$. The resulting
magnetization per spin is
\begin{equation}
m_{\mathrm{F/Ferri}} = 1 - \frac{z}{2(1+z)(e^{h/T} + 1)}.
\end{equation}
In the limit $h \ll T$, this reduces to the previously obtained
zero-field result $m_{\mathrm{F/Ferri}} = (\frac{3z}{4} +
1)/(z+1)$, while for $h \gg T$ it approaches saturation $m = 1$
exponentially.

\paragraph{MD/F, MD/Ferri, and triple point.}

On the MD/F and MD/Ferri critical lines, as well as at the triple
point, the fully polarized state belongs to the ground state
manifold at $h=0$. Specifically, the ferrimagnetic state with
$m=m_{\mathrm{Ferri}}$ applies to the MD/Ferri line. Although an
infinitesimal field would select this polarized state at strictly
zero temperature, the behavior at low but finite temperature is
governed by a competition between entropic and energetic
contributions.

Thus, the partition function is dominated by two competing terms,
$Z\approx Z_{1}+Z_{2}$. $Z_{1}$ originates from exponentially
numerous configurations containing a fraction $1-p_{0}$ of singlet
bonds:
\begin{equation}
Z_{1}\sim \exp \left( \mathcal{S}_{0}\mathcal{N}+\frac{hm_{0}}{T}\mathcal{N}%
\right) .
\end{equation}%
Here $\mathcal{S}_{0}$ is the residual entropy per spin, and
$m_{0}$ is the zero-field magnetization of the ground state
manifold. As shown above, only triangular and cubic lattices at
the triple point exhibit a nonvanishing $m_{0}\approx 0.57$ in the
thermodynamic limit; in all other cases, $m_{0}\to 0$.

$Z_{2}$ arises from the much fewer nearly polarized states above
the percolation threshold, which possess lower Zeeman energy:
\begin{equation*}
Z_{2}\sim \exp \left( \frac{hm_{\max }}{T}\mathcal{N}\right) ,
\end{equation*}%
where $m_{\max }=m_{\mathrm{Ferri}}$ for the MD/Ferri critical
line, and $m_{\max }=1$ for the MD/F critical line and the triple
point.

Comparing $Z_{1}$ and $Z_{2}$, we find that $Z_{1}$ dominates for
$h\ll T$ (weak-field regime), and $Z_{2}$ prevails for $h\gg T$
(strong-field regime). Consequently, in the weak-field regime
$h\ll T$, the system behaves as a paramagnet with $m\sim h/T$ for
all cases except triangular and cubic lattices at the triple
point. For these latter lattices, the ground state consists of two
distinct components: an infinite ferromagnetic cluster
contributing a finite magnetization $m_{0}$, and numerous small
clusters producing a conventional paramagnetic response. The total
magnetization therefore follows $m-m_{0}\sim h/T$.

In the strong-field regime $h\gg T$, the system behaves as a
ferromagnet (or ferrimagnet for the MD/Ferri line), with
magnetization approaching saturation $m=m_{\max }$. For the
MD/Ferri critical line, a further increase of the magnetic field
induces a magnetization jump at $h=h_{1,2}$, which is smoothed by
temperature over a width $\Delta h\sim T$.

Here we should note that for pure Heisenberg model (all $K_{n}=0$)
and higher values of spins $\sigma $, the system is in the MD
phase for $J>2s$, and the magnetization curve has $2\sigma $
magnetization plateaus at values $m_{i}=(s+i)/3$.

For completeness, we note that for the pure Heisenberg model (all
$K_{n}=0$) and higher spin values $\sigma $, the system resides in
the MD phase for $J>2s$, and the magnetization curve exhibits
$2\sigma $ plateaus at values $m_{i}=(s+L)/3$, where $L=0,1,\ldots
2\sigma $.

\section{One-magnon excitations and specific heat}

The one-magnon excitation spectrum above the ground state of the ideal diamond models in
the ground state phases can be calculated exactly. In the
ferromagnetic phase, the spectrum for $S=S_{\text{gs}}-1$ (where
$S_{\text{gs}}$ is the ground state total spin) consists of two
flat bands with energies $E=2(1-J)$ and $E=2$, together with two
dispersive branches. For the square diamond-decorated lattice,
these dispersive branches take the form
\begin{equation}
E(\mathbf{k})=5\pm \sqrt{25-4(2-\cos k_{x}-\cos k_{y})}  \label{Ek}
\end{equation}%
Near $k_{x}=k_{y}=0$, the lower branch exhibits a quadratic
dispersion,
\begin{equation}
E_{2}(\mathbf{k})=\frac{1}{5}\mathbf{k}^{2},  \label{k2}
\end{equation}%
indicating a gapless spectrum characteristic of a conventional
ferromagnet. Analogous calculations for other lattices -
triangular, hexagonal, and cubic - similarly yield a gapless
quadratic low-energy dispersion $E(\mathbf{k})\sim
\frac{1}{z+1}\mathbf{k}^{2}$.

In the ferrimagnetic phase, a gap $\Delta E = 2(J-1)$ separates
the ground state from states with $S = S_{\text{gs}} + 1$, while
gapless excitations with dispersion $E(\mathbf{k}) \sim
\frac{1}{z+2} \mathbf{k}^2$ appear in the $S = S_{\text{gs}} - 1$
sector. In this respect, the properties of the model resemble
those of conventional Lieb-Mattis ferrimagnet. In the MD phase,
states with $S = S_{\text{gs}} + 1$ possess a gap $\Delta E = J -
3K - 2$, whereas the $S = S_{\text{gs}} - 1$ states remain
degenerate with the ground state.

At the triple point, the behavior depends critically on the
lattice geometry. For triangular and cubic lattices, where the
peak $p_{0}$ of the weight function $W(K,N)$ lies above the
percolation threshold $p_{c}$, the dominant ground state
configurations contain infinite ferromagnetic clusters. These
clusters support gapless excitations with quadratic dispersion
$E(\mathbf{k})\sim \mathbf{k}^{2}$. Consequently, the
low-temperature specific heat in these cases is governed by these
gapless modes and takes the form $C\sim T^{d/2}$, where $d$ is the
lattice dimension.

On the MD/F and MD/Ferri phase boundaries, as well as for
hexagonal and square lattices at the triple point, the situation
is markedly different. In these cases, the dominant ground state
configurations consist almost entirely of small, disconnected
ferromagnetic clusters. As established in Sec.~IIB, the
statistical weight of configurations containing an infinite
(spanning) ferromagnetic cluster is exponentially suppressed in
the thermodynamic limit. Consequently, the typical cluster size
remains finite as $N \to \infty$, implying that most local
excitations are gapped.

This structural difference gives rise to a separation of energy
scales in the thermodynamics. Contributions from gapless magnons
of the type described in Eq.~(\ref{k2}) are weighted by the
exponentially small probability of being in a configuration that
supports such long-range excitations. The low-temperature specific
heat therefore results from a competition between two distinct
contributions. The first term, $C_{1}\sim
T^{d/2}e^{-\mathcal{S}_{0}\mathcal{N}}$, arises from the gapless
modes of the rare infinite clusters. The exponential factor with
residual entropy per spin $\mathcal{S}_{0}$ reflects the
suppression of configurations that support such extended
excitations. The second term originates from gapped excitations of
the dominant finite clusters, $C_{2}\sim (\Delta /T)^{2}e^{-\Delta
/T}$, where $\Delta $ is a excitation gap of typical cluster.

Comparing these terms reveals that $C_1$ dominates only at
temperatures satisfying $T < \Delta / (\mathcal{S}_0
\mathcal{N})$. In the thermodynamic limit $N \to \infty$, this
temperature scale vanishes. Therefore, despite the formal presence
of gapless excitations in the spectrum, the low-temperature
thermodynamics of these systems is effectively governed by gapped,
local excitations, resembling that of a paramagnet with a finite
energy gap.

\section{Summary}

In this work, we have studied the ground state properties,
magnetization behavior, and low-temperature thermodynamics of the
ferromagnetic-antiferromagnetic (F-AF) Heisenberg model on
diamond-decorated lattices with ideal diamond units, incorporating
bilinear and higher-order exchange interactions between diagonal
spins. We considered the general case of arbitrary spin values $s$
(monomer spins) and $\sigma $ (diagonal spins). A key feature of
the model is the local conservation of the composite spin on the
diagonal of each diamond, taking values $L=0,1,\dots ,2\sigma $.
The specific case $s=\sigma =1$ with bilinear and biquadratic
interactions was then analyzed in detail as the simplest
nontrivial illustration of the general formalism.

Analysis of the energy levels of a single diamond allowed us to
construct the global phase diagram. For the general case, the
ground state is given by a uniform configuration of composite
spins $L_{i}=L_{gs}$, where $L_{gs}$ minimizes the local diamond
energy. In the pure Heisenberg case, the system undergoes a series
of $2\sigma $ transitions between phases with different optimal
$L$ as the exchange constant $J$ varies. In the general case with
higher-order interactions, there exists a special multicritical
point where the energies of all possible values of composite spin
states are equal, leading to maximal ground state degeneracy.

For the special case $s=\sigma =1$, we construct explicitly the
phase diagram containing three distinct phases: the ferromagnetic
(F) phase ($L=2$), the ferrimagnetic (Ferri) phase ($L=1$), and
the monomer-dimer (MD) phase ($L=0$). All three phases meet at a
triple point, which arises only upon inclusion of the biquadratic
interaction.

On the rigorously determined phase boundaries, the ground state degeneracy
becomes macroscopic, growing exponentially with system size. For the diamond
chain, we derived exact recurrence relations yielding explicit expressions
for the ground state degeneracy on each phase boundary and at the triple
point. For higher-dimensional lattices, the degeneracy problem maps onto a
bond percolation problem, which we solved numerically using a Monte Carlo
approach. While our numerical results are presented for $s=\sigma =1$, the
percolation framework is fully general and can be applied to arbitrary spin
values by appropriately modifying the cluster weight functions. Our results
reveal that the residual entropy per spin depends sensitively on both
lattice geometry and the specific phase boundary, ranging from 30\% to 60\%
of the maximal possible value. The largest residual entropy is consistently
attained at the multicritical point across all lattices studied.

The magnetization behavior exhibits strong dimensionality and
lattice dependence. On the F/Ferri critical line, we derived an
exact analytical expression for the magnetization per spin that is
valid for any lattice. On other critical lines, the magnetization
vanishes in the thermodynamic limit for all lattices except
triangular and cubic lattices at the triple point, where a finite
magnetization per spin $m\approx 0.57$ emerges. This finite
magnetization is attributed to the presence of an infinite
percolation ferromagnetic cluster in the dominant ground state
configurations.

The low-temperature magnetization curves in an external field
reveal a rich structure. In the bulk phases, we identified
characteristic plateaus and jumps associated with the successive
polarization of composite spins. On the critical lines, a
competition between entropically favored disordered configurations
and energetically favored polarized states leads to distinct
scaling regimes. For $h \ll T$, the system behaves as a
paramagnet, while for $h \gg T$, it approaches saturation.

Analysis of the one-magnon excitation spectrum reveals gapless
quadratic dispersions in the F and Ferri phases, characteristic of
conventional ferromagnets and ferrimagnets. In the MD phase the spectrum above macroscopically degenerate ground state is gapped.
On phase boundaries,
where the ground state consists predominantly of non-interacting
finite clusters, the thermodynamics is governed by gapped local
excitations, despite the formal presence of gapless modes. The
contribution of these gapless modes is exponentially suppressed by
the entropic weight of configurations that support extended
clusters, leading to an effective energy gap in the
low-temperature specific heat. For triangular and cubic lattices
at the triple point, however, the specific heat exhibits
ferromagnet-like behavior $C\sim T^{d/2}$ due to the presence of
an infinite ferromagnetic cluster.

The high residual entropy observed in these diamond-decorated spin
systems, reaching up to 60\% of the maximal possible value for the
$s=\sigma =1$ case, holds significant promise for technological
applications, particularly in ultra-low-temperature refrigeration.
Materials with macroscopic ground state degeneracy are ideal
candidates for adiabatic demagnetization cooling, where the large
entropy reservoir enables efficient cooling to millikelvin
temperatures. Beyond conventional refrigeration, these systems are
also well suited for quantum thermal machines, where macroscopic
ground state degeneracy can be harnessed to realize efficient heat
engines and refrigerators operating in the quantum regime.

\bibliography{s1}

\end{document}